\documentclass[10pt]{elsarticle}
\usepackage{lineno,hyperref}
\modulolinenumbers[5]
\usepackage{graphicx}
\usepackage{bm}
\usepackage{subfigure}
\usepackage{mathtools}
\usepackage{amsmath}
\usepackage{epstopdf}
\usepackage{textcomp}
\usepackage{gensymb}
\usepackage{adjustbox}
\usepackage{geometry}
\usepackage{multirow}
\usepackage{braket}
\usepackage{float}
\newcommand{\be}{\begin{eqnarray}}
\newcommand{\ee}{\end{eqnarray}}

\begin{document}
\begin{frontmatter}

\title{Engineering Nonlinear Optical Responses via Inversion Symmetry Breaking in bilayer Bi$_2$Se$_3$}

\author{Vineet Kumar Sharma$^*$}
\address{Department of Physics and Astronomy, Howard University, Washington DC, 20059}
\ead{kvineet66@gmail.com}
\author{Alana Okullo}
\address{Department of Physics and Astronomy, Howard University, Washington DC, 20059}
\author{Barun Ghosh}
\address{Department of Condensed Matter and Materials Physics, S.N. Bose National Center for Basic Sciences, Kolkata, West Bengal 700106}
\author{A. Bansil}
\address{Physics Department, Northeastern University, Boston, MA 02115}
\address{Quantum Materials and Sensing Institute, Northeastern University, Burlington, MA, 01803, USA}
\author{Sugata Chowdhury$^*$}
\address{Department of Physics and Astronomy, Howard University, Washington DC, 20059}
\ead{sugata.chowdhury@howard.edu}

\begin{abstract}
Paucity of naturally occurring noncentrosymmetric materials is stimulating growing interest in engineered two-dimensional systems for nonlinear optical applications. Here, we show that breaking inversion symmetry in centrosymmetric bilayer Bi$_2$Se$_3$ through twisting, point-defect insertion, or the application of an external electric field unlocks rich nonlinear optical responses. In twisted bilayer Bi$_2$Se$_3$ at the first commensurate angle of 21.78$^\circ$, we find peak shift and injection current conductivities of  -14 $ nm.\mu AV^{-2}$ and 104 $\times 10^8$ $nm.A V^{-2}s^{-1}$, respectively, which lie in the visible spectrum and enable efficient THz applications. The external electric field and point-defect insertion both transform the bilayer into C$_ {3v}$ symmetry, with the selenium vacancy (V$_{Se}$) achieving peak shift and injection current conductivities of -190 nm.$\mu AV^{-2}$ and -170 $\times 10^8$ $nm.A V^{-2}s^{-1}$. In all three cases, the peak nonlinear optical responses are found to be comparable to those of benchmark 2D materials such as GeS, and the broadband responses, including helicity-dependent current generation, make these engineered bilayers viable candidates for next-generation 2D photovoltaics.  
\end{abstract}


\end{frontmatter}


\section{Introduction}
Atomically thin two-dimensional (2D) materials are drawing intense interest owing to their tunable optoelectronic properties, which can be controlled through layer count, Moir{\'e} pattern generation, strain application, and defect inclusion. The experimental isolation of graphene in 2004~\cite{graphene2004} opened the door to broader exploration of 2D materials, encompassing transition-metal dichalcogenides (TMDs) such as MoS$_2$ and WS$_2$~\cite{beyondgraphenereview, NonlinearMoS2, NonlinearWS2}, and topological insulators (TIs) such as the  Bi$_2$Se$_3$ family~\cite{chiatti20162d} and MnBi$_2$Te$_4$~\cite{Chen2024}. 

Going beyond the linear optical properties of 2D materials~\cite{deng2018electronic, sahin2012first, chimata2010optical, wang2015nonlinear}, nonlinear light-matter interactions open new frontiers in optoelectronics, driven by the promise of ultrafast nonlinear devices~\cite{thomson2013ultrafast}, entangled photon pair generation~\cite{kwiat1999ultrabright}, and related technologies~\cite{zhao2016atomically, wang2017giant}. Examples of nonlinear optical responses include the generation of nonequilibrium direct current (DC) from periodically driven systems under optical excitation~\cite{oka2009photovoltaic}. Noncentrosymmetric solids exhibit interesting second-order nonlinear DC responses, including shift current (SC) and circular photocurrent (CC)~\cite{wu2017giant, morimoto2016topological, osterhoudt2019colossal, de2017quantized, ma2017direct}, while 2D van der Waals materials more broadly support symmetry-driven second- and third-harmonic generation, making nonlinear optical spectroscopy a powerful characterization tool~\cite{zhao2016atomically, wang2017giant, rangel2017large}. The shift current arises from shifts in charge centers in real space and the associated variations in the Berry connection between the valence and conduction bands~\cite{BPVErecent}. In contrast, the injection current is driven by a Berry curvature asymmetry arising from a change in velocity. Ferroelectric materials such as BiFeO$_3$~\cite{choi2009switchable, yang2010above}, LiNbO$_3$~\cite{glass1974high, fridkin1978anomalous}, BaTiO$_3$~\cite{chynoweth1956surface, koch1975bulk, koch1976anomalous}, and BiTiO$_3$~\cite{choi2009switchable, yang2010above} were among the first found to exhibit the bulk photovoltaic effect (BPVE) in the 1960s and 1970s~\cite{Fridkin2001, bookBPVEncm}. The observation of highly efficient BPVE in these perovskite oxides has sparked interest in exploiting inversion-broken materials with shift current mechanisms as a route to overcoming the Shockley-Queisser limit of conventional p-n junction solar cells.

Nonlinear optical responses can also be induced in centrosymmetric materials through external perturbations such as applied electric fields, as demonstrated in bilayer antiferromagnetic MnBi$_2$Te$_4$~\cite{wang2020electrically}, few-layer WTe$_2$~\cite{xiao2020berry}, and twisted Moir{\'e} systems~\cite{ikeda2020high, liu2020anomalous}, motivating our study of bilayer Bi$_2$Se$_3$-a well-known centrosymmetric topological insulator with a strong Dirac-like surface state~\cite{zhang2009topological} whose linear optical properties have been studied~\cite{tse2015first, deng2018electronic, ermolaev2023broadband} but nonlinear optical responses remain largely unexplored. As a van der Waals material, bilayer Bi$_2$Se$_3$ can be readily engineered to break inversion symmetry, and in this work, we systematically explore the resulting nonlinear optical responses using first-principles calculations. We focus on three distinct symmetry-breaking mechanisms: (i) twisting, (ii) application of an external electric field along the c-axis, and (iii) insertion of point defects. 

\section{Computational Details}
First-principles density functional theory calculations were performed using the augmented plane-wave method as implemented in the Vienna ab-initio Simulation Package (VASP)~\cite{kresse1993ab, kresse1996comput, kresse1996efficiency}. A 350-eV energy cutoff was used for the plane-wave basis set. The force tolerance and energy convergence were set to 10$^{-2}$ and 10$^{-6}$ eV/\AA, respectively. 6$\times$6$\times$1 and 12$\times$ 12$\times$ 1 $\Gamma$-centered k-meshes were used to integrate the Brillouin zone for twisted Moir{\'e} structures and for the other two cases. An effective tight-binding Hamiltonian for localized Bi-$p$ and Se-$p$ orbitals was generated using VASP2WANNIER90~\cite{marzari1997maximally} and Wannier90~\cite{mostofi2014updated} packages. Non-linear current conductivity tensors were computed using WannierBerri~\cite{tsirkin2021high} with a 200 $\times$ 200 $\times$1 k-point grid.  The broadening parameter ($\eta$) used was 0.04 to compute the shift current conductivity tensors. The Twister program~\cite{naik2022twister} was used to generate twisted Moir{\'e} structures at the first commensurate angle of 21.78$^\circ$ for our bilayer systems. Spin-orbit coupling effects were included in all calculations.

\textbf{Theoretical Background:} Second-order nonlinear injection and shift currents under monochromatic light excitation are given by:

\begin{eqnarray}
\frac{dJ_{IC}^{a}}{dt} = -\frac{\pi e^3}{\hbar^2}\int [d\vec{k}] \sum_{n,m, \sigma}f_{nm}\Delta_{mn}^a r_{nm}^b r_{nm}^c \delta(\omega_{mn} - \omega)E^b(\omega)E^c(-\omega)
\end{eqnarray}
\begin{eqnarray}
J_{SC}^{a} = -\frac{i\pi e^3}{2\hbar^2}\int [d\vec{k}] \sum_{n,m, \sigma}f_{nm}(r_{nm}^b r_{nm;k_a}^c - r_{nm}^c r_{nm;k_a}^b)\delta(\omega_{mn} - \omega)E^b(\omega)E^c(-\omega)
\end{eqnarray}

Where a, b, and c  represent the three crystallographic directions, and $r_{nm;k_a}^b$ = $\frac{\delta r_{nm}^b}{\delta k_a}$ - i$r_{nm}^b$ (A$_{m}^a$ - A$_{n}^a$) is a gauge covariant derivative. $\vec{r}_{nm}$ = $\langle n|i\delta_k|m \rangle$ and $A_n$ = $\langle n|i\delta_k|n \rangle$ are the interband and intraband Berry connections, respectively. $f_{nm}$ = $f_{n} - f_{m}$, where $f_m$ and $f_n$ are the Fermi-Dirac distribution functions for the $m^{th}$ and $n^{th}$ bands, respectively. $\hbar \Delta_{mn}$ = $v_{mm}^c - v_{nn}^c$ is the difference in band velocities between states m and n, reflecting the interband velocity asymmetry that drives the injection current. The shift and injection currents can be expressed in terms of the anticommutator $\{r_{nm}^b, r_{nm}^c\} \equiv r_{nm}^b r_{nm}^c + r_{nm}^c r_{nm}^b$ and commutator $[r_{nm}^b, r_{nm}^c] \equiv r_{nm}^cr_{nm}^b - r_{nm}^c r_{nm}^b$ of the position matrix elements. Since the present work focuses on nonmagnetic materials, the full generalized expressions are not reproduced here. The shift and injection current conductivity tensors are given, respectively by:

\begin{eqnarray}
\sigma^{c;ab} = -\frac{\pi e^3}{4 \hbar^2}\int_{k}\sum_{n,m}f_{nm}(R_{n,m}^{c,a} + R_{n,m}^{c,b})\{r_{nm}^b, r_{nm}^a\} \delta(\omega_{mn} - \omega)
\end{eqnarray}

\begin{eqnarray}
\eta^{c;ab} = -\frac{\pi e^3}{2 \hbar^2}\int_{k}\sum_{n,m}f_{nm}^{FD}\Delta_{mn}^c r_{nm}^b r_{nm}^a\delta(\omega_{mn} - \omega)
\end{eqnarray}

\begin{figure*}
\centering
\includegraphics[width=\textwidth]{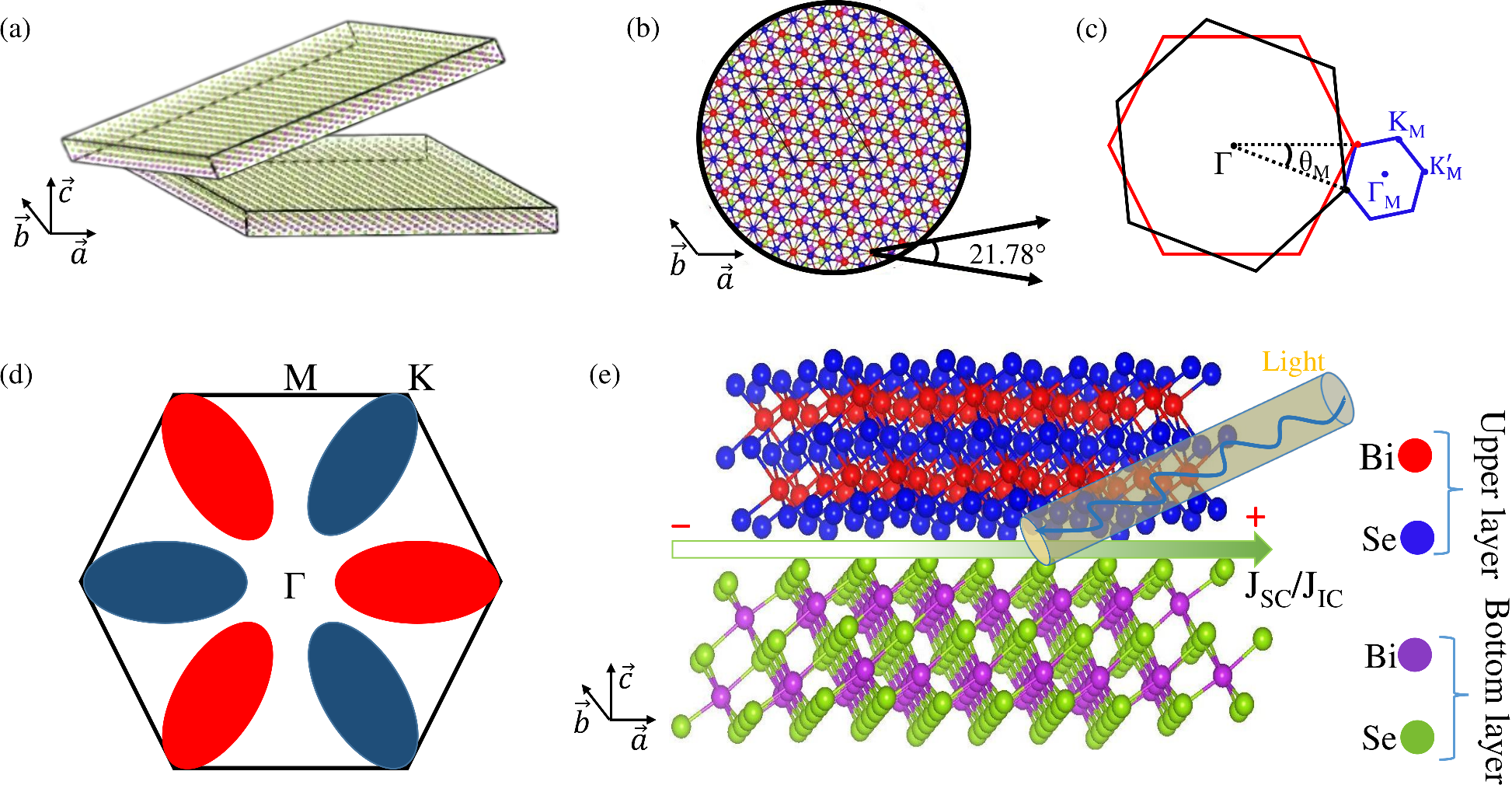}
\caption {(a) Side view of the twisted bilayer system. (b) Top view of the twisted Moir{\'e} pattern with a twist angle of 21.78$^\circ$. (c) 2D Brillouin zone for the twisted Moir{\'e} system. (d) Schematic of a non-zero Berry-curvature dipole induced by non-centrosymmetry. (e) Schematic of the generation of photovoltaic currents in a twisted bilayer.}
\end{figure*}

\section{Results and discussion}
The calculated electronic structure of the bilayer Bi$_2$Se$_3$ is shown in Supplementary Figure S1(a). An analysis of the projected band structure reveals that the valence band is predominantly derived from $Se-p$ orbitals, while both $Bi-p$ and $Se-p$ orbitals contribute to the conduction band [Figure S1(b)]. The semiconducting nature of the band structure is consistent with its potential for optoelectronic applications. 

\subsection{Electronic structure and nonlinear optical response in twisted bilayer Bi$_2$Se$_3$}The twisted Moir{\'e} bilayer was constructed using the commensurate superlattice approach as implemented in the Twister code~\cite{naik2022twister}, which requires identical lattice constants in both layers within the hexagonal family. Commensurate twist angles were obtained using the analytical expression~\cite{lopes2007graphene}:

\begin{eqnarray}
cos (\theta_i) = \frac{n^2 + 4nm + n^2}{2(n^2 + mn + n^2)}
\end{eqnarray} 
Here, n and m are integers, and the corresponding superlattice vectors are A$_1$ = ma$_1$ + na$_2$, A$_2$ = -na$_1$ + (m + n)a$_2$. 

The twisted Moir{\'e} bilayer at the first commensurate angle of 21.78$^{\circ}$ was constructed using Eq. 3.1 with n = 21 and m = -20, yielding superlattice vectors $\sqrt{7}$ times the original lattice vectors and a unit cell containing 70 atoms. Figure 1(a, b) shows a top and side view of the resulting Moir{\'e} pattern. Figure 1(c) depicts the 2D Brillouin zone of the centrosymmetric and twisted Moir{\'e} systems. The symmetry analysis indicates that the twisted system belongs to the non-centrosymmetric point group D3 [space group P312 (\#149)].

The twisted bilayer Bi$_2$Se$_3$ is an indirect bandgap semiconductor with a bandgap of 0.18 eV [Figure 2(a)]. The band structure exhibits Rashba-type splitting near the $\Gamma$-point due to broken inversion symmetry [inset, Figure 2(a)], with orbital contributions around the Fermi level similar to those of the untwisted bilayer [Figure S1(c)]. Unlike bulk Bi$_2$Se$_3$, no band inversion is observed, confirming the trivial semiconducting nature of this system. The broken inversion symmetry gives rise to a non-zero Berry curvature dipole [Figure 2(b)], consistent with the non-centrosymmetry-induced electric polarization between layers [Figure 1(d, e)], which underlies the nonlinear optical response.

The shift and injection current conductivity tensors are shown in Figures 2(c) and 2(d). The absence of optical transitions within the bandgap region is reflected in the conductivity. The three-fold rotational symmetry ($C_{3z}$) of the twisted system yields the equivalent non-zero tensor relations: Xxx = -Xxy = -Xyx = -Xyy. Two non-zero shift current components are identified: $\sigma_{Xxx}$, driven by linearly polarized light along x [inset, Figure 2(c)], and $\sigma_{Xyz}$, generated by light polarized along y and z [Figure 2(c)], where the capital letter denotes the photocurrent direction and lowercase letters denote the electric field polarization directions at frequencies $\omega$ and $-\omega$. The peak values of $\sigma_{Xyz}$ and $\sigma_{Xxx}$ are -14 nm.$\mu AV^{-2}$ and -11 $\times$10$^{-3}$ nm.$\mu AV^{-2}$, respectively, both occurring near 3 eV. The injection current conductivity tensor $\eta_{Xyz}$ reaches a maximum of 104 $\times$10$^8$ $nm.A V^{-2}s^{-1}$ near 1.7 eV [Figure 2(d)]. We also consider $\eta_{Xzy}$ to illustrate the circular photogalvanic effects in Figure 2(d), which shows that $\eta_{Xyz}$ = -$\eta_{Xzy}$, so that $\eta_{Xyz}$ - $\eta_{Xzy}$ = 2$\eta_{Xyz}$. This implies that the current reverses sign under switching between left- and right-circularly polarized light, demonstrating that a net photocurrent can be generated without an external bias.

\begin{figure*}
\includegraphics[width=\textwidth]{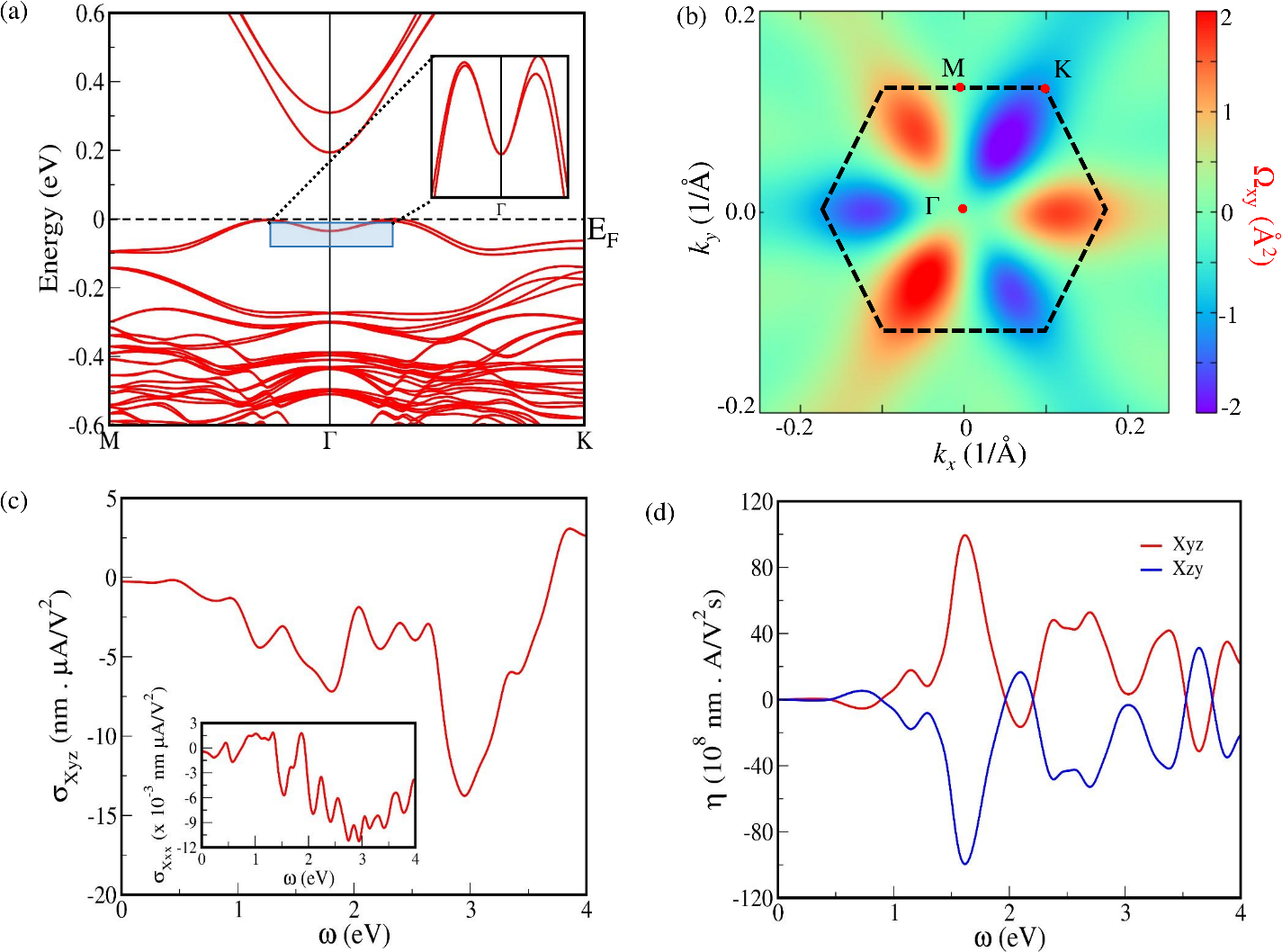}
\caption{(a) Electronic band structure of twisted bilayer Bi$_2$Se$_3$, with the inset showing Rashba-type band splitting at the $\Gamma$-point induced by the broken inversion symmetry. (b) Berry curvature $\Omega_{xy}$, demonstrating a non-zero Berry curvature dipole. (c) Shift-current and (d) injection-current conductivity tensors, reflecting the linear and circular photogalvanic effects, respectively.}
\end{figure*}

\begin{figure*}
\centering
\includegraphics[width=\textwidth]{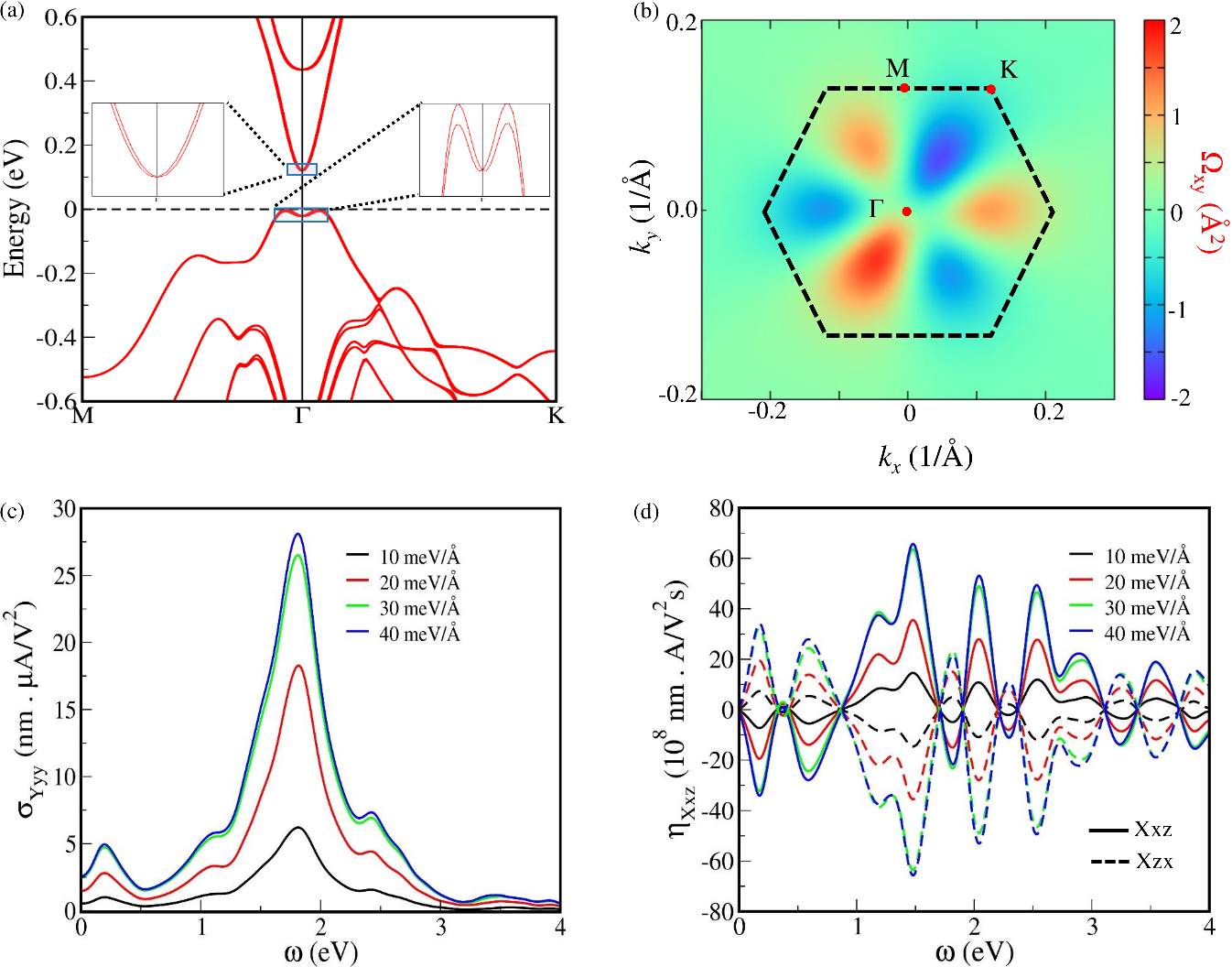}
\caption {Electronic structure and nonlinear optical properties of bilayer Bi$_2$Se$_3$ under an external electric field. (a) Band structure at 40 meV/$\AA$, with inset showing Rashba-type band splitting near the $\Gamma$-point induced by broken inversion symmetry. (b) Berry curvature $\Omega_{xy}$ demonstrating a non-zero Berry curvature dipole. (c) Shift current and (d) injection current conductivity at various electric field strengths.}
\end{figure*}

\subsection{External electric field effects on bilayer Bi$_2$Se$_3$}
Recent studies on electrically gated bilayer MnBi$_2$Te$_4$~\cite{wang2020electrically} show that an external electric field (E$_z$) can break inversion symmetry, transforming a centrosymmetric bilayer into a non-centrosymmetric system with space group P3m1 (\#156). Figure 3 displays the calculated electronic and nonlinear optical properties of the Bi$_2$Se$_3$ bilayer under several different values of an external electric field (E$_z$).  The band structure at 40 meV/$\AA$ [Figure 3(a)] confirms the small bandgap semiconducting character of the system, with bands around the Fermi level dominated by $Bi-p$ and $Se-p$ orbitals, consistent with the untwisted and twisted bilayer cases [Supplementary Figure S2(a)]. Electronic responses at different electric field strengths [Supplementary Figure S2(b–d)] show Rashba-type splitting in all cases [insets, Figures 3(a) and S2(b–d)], and non-zero Berry curvature dipole calculations confirm the non-centrosymmetric character of these nonmagnetic systems.

Figure 3(c) shows the computed shift current conductivity for y-polarized light. Symmetry analysis under C$_{3z}$ yields the relations Yyy = -Xxy = -Xyx = -Xyy, so that we have three independent shift-current conductivity tensor components: $\sigma_{Yyy}$, $\sigma_{Yyz}$, and $\sigma_{Zxx}$. The maximum shift-current conductivity $\sigma_{Yyy}$ reaches 5.4 nm.$\mu AV^{-2}$ at a photon energy of $\sim$1.7 eV for an applied field of 10 meV/$\AA$, increasing substantially with field strength to 28 nm.$\mu AV^{-2}$ at 40 meV/$\AA$. In addition to the nonlinear optical response in the visible spectrum, a significant signal is found in the infrared and THz regimes, supported by the first interband transition between the valence band maximum and conduction band minimum [Figure 3(a), Figure S2(b-d)]. The peak shift-current conductivities of $\sigma_{Yyz}$ and $\sigma_{Zxx}$ reach 15 nm.$\mu AV^{-2}$ and 27 nm.$\mu AV^{-2}$, respectively, at 40 meV/$\AA$ [Figure S3(a,b)], while $\sigma_{Xyz}$ and $\sigma_{Xxx}$ are absent due to mirror symmetry ($\overline{M_x}$). The shift current conductivity tensor at -40 meV/$\AA$, alongside that at +40 meV/Å [Figure S3(c)], demonstrates a sign reversal upon switching the field direction, consistent with the polar nature of the bilayer under an applied electric field.

There is a single non-zero normal injection current (NIC) tensor component $\eta_{Xxz}$ in the electrically gated bilayer. Two strong peaks are found in the infrared region, with NIC conductivity values of 12.6 $\times10^8$ nm.$AV^ {-2}s^{-1}$ and 9 $\times$10$^8$ nm.$AV^ {-2}s^{-1}$ between 1.4 and 2.2 eV [Figure 3(d)], reaching 64$\times$10$^8$ nm.$AV^ {-2}s^{-1}$ at 40 meV/$\AA$. Note that this $\eta_{Xxz}$ is unchanged upon reversal of the electric field direction. The relation $\eta_{Xxz}$ = -$\eta_{Xzx}$ confirms helicity-dependent current generation, consistent with the twisted bilayer case. The strong nonlinear optical response in the THz regime makes electrically gated bilayer Bi$_2$Se$_3$ a promising candidate for future low-dimensional optoelectronic applications.

\subsection{Point-defect-driven nonlinear optical responses}
We now turn to consider the effects of point defects that break the inversion symmetry of the Bi$_2$Se$_3$ bilayer. Three different point defects are considered: selenium vacancies (V$_{Se}$), Se$_{Bi}$ antisite defects, and Bi$_{Se}$ antisite defects [Figure 4(a, b, c)], all studied using a 2×2×1 supercell. The V$_{Se}$ and Se$_{Bi}$ cases introduce 4.16\% defects by removing or substituting one Se atom [Figures 4(a,b)], while the Bi$_{Se}$ case introduces 6.25\% defects by substituting one Bi atom with Se. Each defect is introduced in one quintuple layer, breaking inversion symmetry and yielding a non-centrosymmetric bilayer with space group P3m1 (\#156). The calculated electronic structures [Figures 4(d, e, f)] reveal a semiconducting-to-metallic transition in all three cases. The symmetry analogy with the electrically gated bilayer, together with Rashba-type splitting at the $\Gamma$-point, confirms a non-zero Berry curvature dipole driven by the broken inversion symmetry. The metallic nature of defect-induced systems is not a barrier to nonlinear optical responses, as established by extensive studies of shift and injection currents in Weyl and Dirac semimetals~\cite{ahn2020low, roy2025large}.

\begin{figure*}
\centering
\includegraphics[width=\textwidth]{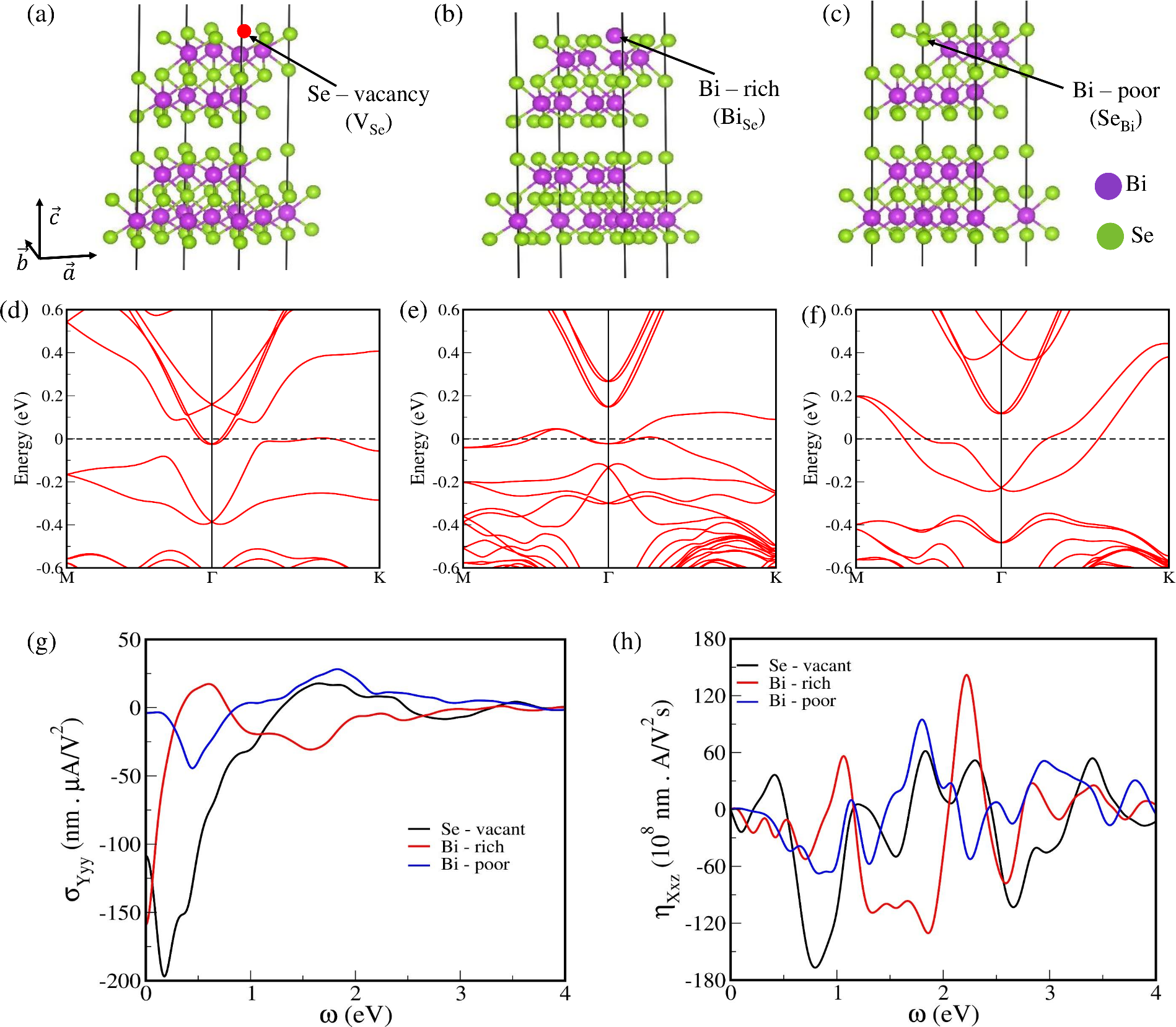}
\caption {Point defects in bilayer Bi$_2$Se$_3$ and nonlinear responses: (a) Se vacancy defect (red circle). (b) Bi-rich antisite defect induced by replacing Se with Bi. (c) Bi-poor antisite defect induced by replacing Bi with Se. (d-f) Band structures corresponding to the defects in panels (a-c). (g) Shift current and (h) injection current conductivities.}
\end{figure*}

Figure 4(g) displays $\sigma_{Yyy}$, one of three independent non-zero shift current conductivity tensor components in these bilayers. The peak shift current conductivity reaches -190 nm.$\mu AV^{-2}$ for the V$_{Se}$ case near 0.2 eV, while the Bi$_{Se}$ and Se$_{Bi}$ cases yield maxima of -40 nm.$\mu AV^{-2}$ and -47 nm.$\mu AV^{-2}$ near 1.6 eV and 0.4 eV, respectively, indicating a dominant nonlinear response in the infrared region. The remaining non-zero components $\sigma_{Yyz}$ and $\sigma_{Zxx}$ are shown in Figure S4; notably, their peak values are lower by an order of magnitude for V$_{Se}$, while Bi$_{Se}$ and Se$_{Bi}$ exhibit peak conductivities in $\sigma_{Yyz}$ and $\sigma_{Zxx}$ comparable to the V$_{Se}$ case in $\sigma_{Yyy}$. The injection current conductivity tensor component $\eta_{Xxz}$ [Figure 4(h)] shows multiple peaks across a broad energy range, likely reflecting optical transitions from the topmost valence band to multiple conduction band states enabled by the substantial band splitting. As in the twisted and electrically gated bilayer cases, $\eta_{Xxz}$ = -$\eta_{Xzx}$, confirming helicity-dependent current generation. Among the defect cases examined, V$_{Se}$ yields the strongest nonlinear optical responses, establishing point defect engineering as a viable route to nonlinear optical responses in 2D materials. To place these findings in context, Table 1 compares our computed peak conductivities with those of well-known 2D systems

\begin{table*}
\caption{Comparative studies of calculated nonlinear optical conductivities with well-known bulk and 2D non-centrosymmetric materials}
\centering
\begin{tabular}{|c|c|c|c|}
\hline
Materials & $\sigma ( nm.\mu AV^{-2})$ & $\eta (10^8 nm.AV^ {-2}s^{-1})$ & Reference\\
\hline
Twisted bilayer Bi$_2$Se$_3$ & 14 & 104 & This work \\
\hline
Bilayer Bi$_2$Se$_3$ at 40 meV/$\AA$ & 28 & 64 & This work \\
\hline
Bi$_2$Se$_3$ for V$_{Se}$ & 190 & 170 & This work \\
\hline
GeS (2D)&100&100-1000&~\cite{panday2019injection}\\
\hline
MoS$_2$ (2D)&--&10$^{-7}$&~\cite{arzate2016optical}\\
\hline
\end{tabular}
\end{table*}

\section{Summary}
We discuss the nonlinear optical responses arising from systematic inversion symmetry breaking in centrosymmetric van der Waals bilayer Bi$_2$Se$_3$, achieved through three distinct mechanisms: twisting, application of an external electric field, and insertion of point defects. The twisted Moir{\'e} bilayer exhibits significant nonlinear optical responses with predominant injection current conductivity, including helicity-dependent current generation. Since these responses are symmetry-constrained, the electric field and point-defect cases produce polarization-dependent responses that are qualitatively distinct from those of the Moir{\'e} bilayer; notably, both exhibit large nonlinear responses and sign switching in the injection current under helicity reversal. Additionally, reversing the external electric field switches the sign of the shift current. Across all three symmetry-breaking mechanisms, peak nonlinear optical conductivities under polarized light make these engineered non-centrosymmetric bilayers viable candidates for next-generation 2D photovoltaics.

\section{Acknowledgements}
V.K.S., S.C., and A.B. acknowledge the National Science Foundation through the Expand-QISE award NSF-OMA-2329067 for financial support. A.O. thanks the IBM-HBCU Quantum Center for financial support. The research at Howard University used the resources of Accelerate ACCESS PHYS220127 and PHYS2100073. The work at Howard University also benefited from the resources of Northeastern University’s Advanced Scientific Computation Center, the Discovery Cluster. B. G. acknowledges the Prime Minister Early Career Research Grant (PM-ECRG) from Anusandhan National Research Foundation (ANRF), file number ANRF/ECRG/2024/003677/PMS.


\end{document}